\documentclass[aps,prl,twocolumn,superscriptaddress,showpacs,amsmath,amssymb]{revtex4-1}
\usepackage{amsmath}
\usepackage{amsfonts}
\usepackage{amssymb}
\usepackage{graphicx}
\usepackage{color}

\usepackage{bm}
\usepackage{braket}
\usepackage{hyperref}

\begin{document}

\title{Electric probe for the toric code phase in Kitaev materials \\
through the hyperfine interaction}

\author{Masahiko G. Yamada}
\email[]{myamada@mp.es.osaka-u.ac.jp}
\affiliation{Department of Materials Engineering Science, Osaka University, Toyonaka 560-8531, Japan}
\author{Satoshi Fujimoto}
\affiliation{Department of Materials Engineering Science, Osaka University, Toyonaka 560-8531, Japan}
\affiliation{Center for Quantum Information and Quantum Biology, Osaka University, Toyonaka 560-8531, Japan}

\date{\today}

\begin{abstract}
The Kitaev model is a remarkable spin model with gapped and
gapless spin liquid phases, which are potentially realized in
iridates and $\alpha$-RuCl$_3$.  In the recent experiment of
$\alpha$-RuCl$_3$, the signature of a nematic transition to the gapped toric
code phase, which breaks the $C_3$ symmetry of the system,
has been observed through the angle dependence of the heat capacity.
We here propose a mechanism by which the nematic transition can
be detected electrically.  This is seemingly impossible because
$J_\textrm{eff}=1/2$ spins do not have an electric quadrupole
moment (EQM).  However, in the second-order perturbation the
virtual state with a nonzero EQM appears, which makes the
nematic order parameter detectable by nuclear magnetic resonance
and M\"ossbauer spectroscopy.  The purely magnetic origin of EQM
is different from conventional electronic nematic phases, allowing
the direct detection of the realization of Kitaev's toric
error-correction code.
\end{abstract}

\maketitle

\textit{Introduction}.---
The Kitaev model~\cite{Kitaev2006} is a notable spin model
for quantum spin liquids (QSLs) with gapped and
gapless ground states.  After pioneering work by
Jackeli and Khaliullin~\cite{Jackeli2009}, potential experimental
realizations were reported in iridates~\cite{Singh2010,Singh2012}
and $\alpha$-RuCl$_3$~\cite{Plumb2014}.  Indeed, those
materials have $d^5$ metal ions in the octahedral ligand field forming
the honeycomb lattice, which results in unusual anisotropic
interactions proposed by Kitaev~\cite{Kitaev2006}.
This Jackeli-Khaliullin mechanism is intrinsic to the
$J_\textrm{eff}=1/2$ magnetic moment with a strong spin-orbit
coupling (SOC), and makes the $d^5$ materials family,
sometimes called Kitaev materials, a fascinating platform
for the physics of Majorana fermions.  Especially, after
the discovery of a field-revealed QSL phase in
$\alpha$-RuCl$_3$~\cite{KasaharaSugii2018,Kasahara2018},
various experimental techniques were used to characterize
this exotic phase under a magnetic field~\cite{Banerjee2018,Jansa2018,Lebert2020}.
However, the realization of Kitaev's gapped $A$ phase,
which is nothing but a toric code phase~\cite{Kitaev2003}, was
only discussed in a complex structure in metal-organic
frameworks~\cite{Yamada2017MOF}.

Kitaev's $A$ phase is the ground state of the Kitaev model
in the anisotropic limit.  This is a gapped $Z_2$ spin liquid
phase and is mapped to the toric code model in the fourth
order perturbation.  The toric code is a topological error
correction code which is useful in fault-tolerant quantum computing.
We here discuss another route towards the realization of
this phase.  This toric code phase is potentially
realized by a spontaneous breaking of the $C_3$ symmetry of
the isotropic Kitaev model.  If the order parameter reaches
a critical value, the system transforms from $B$ phase
to $A$ phase.  This order parameter consists of
quadrupole operators, rather than usual magnetic dipoles,
and in this sense we can regard it as a nematic transition.

On the analogy of liquid crystals, a nematic
phase is discussed in various fields of condensed matter physics,
ranging from spin nematic phases in frustrated magnets~\cite{Penc2011}
to electronic nematic phases in quantum Hall systems~\cite{Lilly1999},
ruthanates~\cite{Borzi2007}, unconventional
superconductors~\cite{delaCruz2008}, \textit{etc}.
Inspired by the previous numerical studies~\cite{Gordon2019,Lee2020},
we seek for a possibility of the nematic transition in Kitaev
materials.  In $J_\textrm{eff}=1/2$ Kitaev materials, it should
be called \textit{spin-orbital nematic}~\cite{Li2020} with properties
of both spin nematic and electronic nematic.

Recently, O.~Tanaka \textit{et al.}~\cite{Tanaka2020} indeed observed
such a spin-orbital nematic transition from a gapped chiral spin liquid
phase to a different gapped phase characterized by the broken
threefold rotation ($C_3$) symmetry, based on the measurements
of the angle dependence of heat capacity under a strong magnetic field.
It has been proposed that this symmetry-broken phase could be the
toric code phase~\cite{Takahashi2021}, as the half-quantized thermal
Hall effect disappears at the transition point~\cite{Kasahara2018}.
However, the property of this nematic transition is still obscure,
and we need a more sensitive local probe for this unusual phase transition.

Therefore, we propose an electric quadrupole moment (EQM) as
a direct probe for the topological nematic transition~\cite{Takahashi2021}
of the $J_\textrm{eff}=1/2$ magnetic moments.  This statement is very
counterintuitive as the $J_\textrm{eff}=1/2$ pseudospin does
not have an EQM in the cubic environment, differently from
the $J_\textrm{eff}=3/2$ case~\cite{Yamada2018},
where the quadrupole moment is directly measurable.  Interestingly,
however, holes with a $J_\textrm{eff}=1/2$ pseudospin can hop to the
nearest-neighbor (NN) sites, and an virtual state with two holes
can possess an EQM.  This is because via the superexchange pathway
involving the Cl $p$-orbitals the $J_\textrm{eff}=1/2$ state can
be transformed into a state with a nonzero quadrupole moment.
This enables us to \textit{electrically} detect the nematic
order parameter, which is originally written in terms of
spin operators.  We also discuss
that, although the Chern number is not measurable, its change
can be inferred from the careful analysis of the derivative of
the in-plane anisotropy parameter $\eta$.

In a real experimental setup, the most sensitive way to
measure the EQM is through the hyperfine interaction because
the nuclear with a spin $I \geq 1$ can feel the electric
field gradient (EFG), or the EQM.
Especially, nuclear magnetic resonance (NMR) and
M\"ossbauer spectroscopy (MS) use a nuclear spin of Ru
as a direct probe, and they are highly sensitive to the symmetry of
the local environment.
If the $C_3$ symmetry of Ru forming the honeycomb lattice is broken,
it can potentially be detected by $^{99/101}$Ru-NMR~\cite{Majumder2015},
or $^{99}$Ru-MS~\cite{Kobayashi1992}.  In NMR and MS,
the in-plane anisotropy is characterized by a single
dimensionless parameter $\eta$~\cite{Toyoda2018nematic,Toyoda2018inplane,Kitagawa2018}.
If the EFG or EQM tensor has an anisotropy around the [111] axis,
$\eta$ gets nonzero and the signal splits or shifts,
which could detect the existence of a nematic order.

In this Letter, we will prove that the in-plane anisotropy $\eta$
is directly connected to the nematic
order parameter in terms of Majorana fermions, which potentially
detects the transition to the toric code phase.

\begin{figure}
\centering
\includegraphics[width=8.6cm]{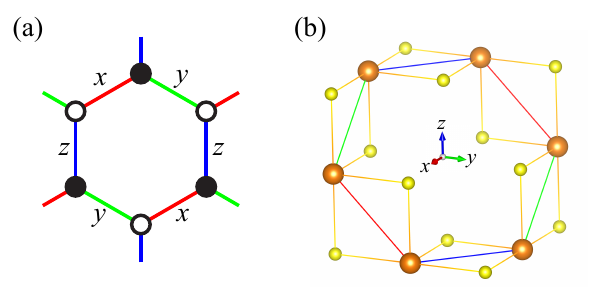}
\caption{(a) Honeycomb lattice where the Kitaev model is defined.
Red, green, and blue bonds represent bonds in the $x$-, $y$-, and
$z$-directions, respectively.
(b) Idealized geometry of $\alpha$-RuCl$_3$.  Orange and yellow
spheres represent Ru and Cl ions, respectively.  Bonds in the
$\gamma$-direction are defined to be perpendicular to the $\gamma$-axis
of the cubic lattice.  The figure is generated by VESTA~\cite{Momma2011}.}
\label{honeycomb}
\end{figure}

\textit{Quadrupole moment}.---
An electronic EQM is defined for $d$-orbitals by
\begin{equation}
    q^{\alpha\beta} = \frac{3}{2}(L^\alpha L^\beta + L^\beta L^\alpha) - \bm{L}^2\delta^{\alpha\beta},
\end{equation}
where $\bm{L}$ are $L=2$ orbital angular momentum operators
of Ru $d$-orbitals and $\alpha=x$, $y$, or $z$, and $\beta=x$, $y$,
or $z$.  This rank-2 traceless symmetric tensor directly couples
to the nuclear EQM of Ru, and the anisotropy of $q^{\alpha\beta}$
is easily measurable.  If the EFG from the surrounding ions is
negligible as is the case for $^{99}$Ru-MS~\cite{Kobayashi1992},
we can identify the effective EFG $V_\textrm{eff}^{\alpha\beta}$
to be proportional to $q^{\alpha\beta}$.  Therefore, we will not
distinguish between EFG and EQM of Ru from now on.

The definition of $\eta$ in terms of $q^{\alpha\beta}$ is as follows.
Since this tensor is symmetric, it can be diagonalized by orthogonal
transformation.  Here we denote the principal axis as $abc$, where
we define the order of $abc$ such that $|q^{cc}| \geq |q^{bb}| \geq |q^{aa}|$.
In this case, $\eta$ is defined as $\eta = (q^{aa} - q^{bb}) / q^{cc}$.
If $\eta = 0$, it is apparent that EQM is invariant under the rotation
around the $c$-axis, and thus it potentially detects the breaking
of the $C_3$ symmetry of $\alpha$-RuCl$_3$.  However, the connection
between $\eta$ and the nematic order parameter is not evident
in this form.  Differently from the ``electronic'' nematic order,
where $\eta$ detects the distortion of surrounding ligands,
the spin nematic order is subtle without a detectable
structural transition.

Since the nematic transition of $\alpha$-RuCl$_3$ may be purely
magnetic as around the transition point $H \sim 10$ T no
structural transition has been observed~\cite{Tanaka2020},
we have to think of a mechanism where a pure spin
operator is transformed into an electric quadrupole.
Especially, in the case where the position of Cl ligands is not
distorted, we have to consider a purely electronic origin
for this mechanism, which involves a microscopic structure
of Ru $d$-orbitals.  From now on we set $\hbar=1$.

As is well-known, the $J_\textrm{eff}=1/2$ pseudospin cannot possess
an EQM in the cubic environment, thus we have to perturb the
$J_\textrm{eff}=1/2$ wavefunction in some way to get a nonzero
expectation value of EQM.  One simple way is by the ligand field
effect of the lattice distortion, but it only produces a static
contribution.  A more exotic answer is to perturb the
$J_\textrm{eff}=1/2$ wavefunction via the superexchange mechanism.
Especially, in the case of the low-spin $d^5$ configuration,
it is well-known as the Jackeli-Khaliullin mechanism that the
$J_\textrm{eff}=1/2$ state is transformed into $J_\textrm{eff}=3/2$
state with a nonzero quadrupole moment, which produces the following
Kitaev Hamiltonian for $J_\textrm{eff}=1/2$ pseudospins:
\begin{equation}
    \mathcal{H}_\textrm{Kitaev}=-K\sum_{\langle ij\rangle \in \gamma} S_i^\gamma S_j^\gamma,
\end{equation}
where $\bm{S}_i$ is a pseudospin on the $i$th site of $\alpha$-RuCl$_3$,
$K>0$ is a Kitaev interaction, and $\langle ij\rangle \in \gamma$
means an NN bond $\langle ij\rangle$ in the $\gamma$-direction with
$\gamma=x$, $y$, and $z$.
The bond direction is defined as illustrated in Fig.~\ref{honeycomb}(a).
Assuming the 0-flux ground state, the Hamiltonian can be recast into
the tight-binding model of Majorana fermions.
\begin{equation}
    \mathcal{H}_\textrm{Majorana}=\frac{K}{4}\sum_{\langle ij\rangle} ic_i c_j,
\end{equation}
where $c_i$ is an itinerant Majorana fermion on the $i$th site.
We note that in this Letter we do not antisymmetrize Majorana
fermion operators.

Similarly to the Jackeli-Khaliullin mechanism, we can compute an
effective quadrupole moment produced by the virtual state,
and it can potentially have a form of
$S_i^\gamma S_j^\gamma$.  This is how the pure spin operator
$S_i^\gamma S_j^\gamma$ can be transformed into an electric quadrupole
$q^{\gamma\gamma}$ in the second-order perturbation.

\textit{Second-order perturbation}.---
Following Jackeli and Khaliullin~\cite{Jackeli2009}, we will
do the perturbation inside the $t_{2g}$-orbitals assuming a
large octahedral ligand field.  The discussion also follows
Refs.~\cite{Bolens2018optical,Bolens2018me,Pereira2020}.
Especially, the idea is related to the one
discussed in Ref.~\cite{Bulaevskii2008}.
We first note that $t_{2g}$-orbitals
($d_{yz}$, $d_{xz}$, and $d_{xy}$) possess an effective angular
momentum operator $\bm{l}_\textrm{eff}$ with $l_\textrm{eff}=1$.
This effective moment has a relation $\bm{L}=-\bm{l}_\textrm{eff}$
inside the $t_{2g}$-manifold, but we cannot simply use this relation
in the calculation of $q^{\alpha\beta}$.  The computation of
$q^{\alpha\beta}$ involves intermediate $e_g$-orbitals, which
brings about a nonzero correction.  Details are included in
Supplemental Material (SM)~\cite{SM}.

We take the following basis set to write down the Hamiltonian:
\begin{equation}
    \bm{d}_i^\dagger = \left(d_{i,yz,\uparrow}^\dagger, d_{i,yz,\downarrow}^\dagger, d_{i,xz,\uparrow}^\dagger, d_{i,xz,\downarrow}^\dagger, d_{i,xy,\uparrow}^\dagger, d_{i,xy,\downarrow}^\dagger \right),
\end{equation}
where $d_{i,\alpha,\sigma}^\dagger$ denotes a hole creation operator
for a $d_\alpha$-orbital with a spin $\sigma=\uparrow$, $\downarrow$
with $\alpha = yz$, $xz$, and $xy$.  We sometimes identify $yz$, $xz$,
and $xy$ with $x$, $y$, and $z$, respectively.

The Hamiltonian $\mathcal{H}$ consists of the following terms:
\begin{equation}
    \mathcal{H}=\mathcal{H}_\textrm{hop} + \mathcal{H}_\textrm{SOC} + \mathcal{H}_\textrm{LF} + \mathcal{H}_\textrm{Hubbard},
\end{equation}
which is the sum of the kinetic hopping term, the SOC, the ligand
field splitting, and the Hubbard term.  The kinetic hopping term
can be written generically as follows:
\begin{equation}
    \mathcal{H}_\textrm{hop} = -\sum_{\langle ij \rangle \in \gamma} \left[\bm{d}_i^\dagger(T^\gamma \otimes \openone_2)\bm{d}_j + \textrm{H.c.}\right],
\end{equation}
where $\openone_n$ is the $n\times n$ identity matrix, and $T^\gamma$
with $\gamma=x$, $y$, and $z$ are
\begin{align}
    T^x &= \begin{pmatrix}
        0 & 0 & 0 \\
        0 & 0 & t_2 \\
        0 & t_2 & 0
    \end{pmatrix}, \quad
    T^y = \begin{pmatrix}
        0 & 0 & t_2 \\
        0 & 0 & 0 \\
        t_2 & 0 & 0
    \end{pmatrix}, \nonumber \\
    T^z &= \begin{pmatrix}
        0 & t_2 & 0 \\
        t_2 & 0 & 0 \\
        0 & 0 & 0
    \end{pmatrix},
\end{align}
where $t_2$ is the main contribution coming from
the pathway via Cl $p$-orbitals. Of course, we can consider a more generic
form including $t_i$ ($i=1,\dots,4$)~\cite{Bolens2018optical,Bolens2018me}.

The SOC Hamiltonian is $\mathcal{H}_\textrm{SOC} = (\lambda/2)\sum_{i,\alpha} \bm{d}_i^\dagger (l^\alpha \otimes \sigma^\alpha) \bm{d}_i$,
where $\lambda>0$, $(l^\alpha)_{\beta\gamma}=-i\epsilon_{\alpha\beta\gamma}$,
and $\sigma^\alpha$ are Pauli matrices with $\alpha = x$, $y$, and $z$.
$\mathcal{H}_\textrm{LF} = \Delta\sum_i \bm{d}_i^\dagger \left[(\bm{l}\cdot\hat{\bm{n}})^2\otimes \openone_2\right] \bm{d}_i$
with $\hat{\bm{n}}=(1,1,1)/\sqrt{3}$, assuming the preserved $C_3$
symmetry of the lattice.

$\mathcal{H}_\textrm{Hubbard}$ is a multiorbital Hubbard interaction
term.  We here ignore the Hund coupling $J_H$ for simplicity as
$J_H$ is much smaller than the Hubbard interaction $U$.
$\mathcal{H}_\textrm{Hubbard} = (U/2)\sum_i{n_i(n_i - 1)}$,
where $n_i=\bm{d}_i^\dagger \cdot \bm{d}_i$ is a number operator
for each site.

Let us begin with the case without a ligand field splitting by
setting $\Delta=0$.  In the atomic limit without a kinetic term,
the system has exactly one hole per site.  The states for a single hole
are split into $J_\textrm{eff}=3/2$ and $J_\textrm{eff}=1/2$, and
the atomic ground state consists of degenerate $J_\textrm{eff}=1/2$
pseudospins as $\lambda>0$, which is denoted by $\bm{S}_i$.
The effective operator form of $q^{\alpha\beta}$ in
terms of pseudospins $\bm{S}_i$ can be derived from the
second-order perturbation in the kinetic term.
This is achieved by perturbing a magnetic state $\ket{\phi_m}$
into $\ket{\psi_m}$ up to the first order and by computing
\begin{equation}
    \left[q_\textrm{eff}^{\alpha\beta}\right]_{mn} = \braket{\psi_m|q^{\alpha\beta}|\psi_n}.
\end{equation}
$\ket{\psi_m}$ is
\begin{equation}
    \ket{\psi_m} = \alpha\ket{\phi_m} + \frac{1-P}{E_0-\mathcal{H}_0}\mathcal{H}_\textrm{hop}\ket{\phi_m},
\end{equation}
where $\alpha \sim 1$ is a renormalization constant,
$P$ is a projection operator onto unperturbed states, and
$\mathcal{H}_0$ is an unperturbed Hamiltonian with an energy
$E_0$ for $\ket{\phi_m}$.  Since the original $J_\textrm{eff}=1/2$
state $\ket{\phi_m}$ does not have an EQM, the effective operator
can finally be written
\begin{equation}
    q_\textrm{eff}^{\alpha\beta} = P\mathcal{H}_\textrm{hop}\frac{1-P}{E_0-\mathcal{H}_0}q^{\alpha\beta}\frac{1-P}{E_0-\mathcal{H}_0}\mathcal{H}_\textrm{hop}P.
\end{equation}
The contribution of the $\langle ij \rangle$ bond to the $i$th
site can also be written as
\begin{equation}
    q_{ij}^{\alpha\beta} = P\mathcal{H}_\textrm{hop}^{i\rightarrow j}\frac{1-P}{E_0-\mathcal{H}_0}q_i^{\alpha\beta}\frac{1-P}{E_0-\mathcal{H}_0}\mathcal{H}_\textrm{hop}^{j\rightarrow i}P,
\end{equation}
where $\mathcal{H}_\textrm{hop}^{j\rightarrow i} = \bm{d}_i^\dagger(T^\gamma \otimes \openone_2)\bm{d}_j$
when $\langle ij \rangle \in \gamma$.

From now on, an NN site of $i$ is denoted by $i_\gamma$ for the
$\gamma$-direction.  When $\gamma=z$, the direct calculation
leads to the following effective EQM:
\begin{widetext}
\begin{equation}
    q_{ii_z} = \frac{t_2^2}{(U + \frac{3}{2}\lambda)^2}\begin{pmatrix}
\frac{4}{3}(S_i^x S_{i_z}^x - S_i^y S_{i_z}^y) + 4S_i^z S_{i_z}^z & -\frac{4}{3}(S_i^x S_{i_z}^y + S_i^y S_{i_z}^x) & -\frac{16}{3}S_i^x S_{i_z}^z \\
-\frac{4}{3}(S_i^x S_{i_z}^y + S_i^y S_{i_z}^x) & \frac{4}{3}(S_i^y S_{i_z}^y - S_i^x S_{i_z}^x) + 4S_i^z S_{i_z}^z & -\frac{16}{3}S_i^y S_{i_z}^z \\
-\frac{16}{3}S_i^x S_{i_z}^z & -\frac{16}{3}S_i^y S_{i_z}^z & -8S_i^z S_{i_z}^z
\end{pmatrix},\label{zbond}
\end{equation}
\end{widetext}
up to a trivial constant.
Though it looks complicated, the main contribution is simple.
In the spirit of Kitaev's perturbative treatment of the magnetic field,
we can regard the first contribution to be the one which does not change
the flux sector.  In $q_{ii_z}^{\alpha\beta}$, such a contribution is
only the $S_i^z S_{i_z}^z$ term in the diagonal element, which can
be written, assuming that $i$ is on the even sublattice, as
\begin{equation}
    P_0 q_{ii_z} P_0 = \frac{t_2^2}{(U + \frac{3}{2}\lambda)^2}\begin{pmatrix}
-ic_i c_{i_z} & 0 & 0 \\
0 & -ic_i c_{i_z} & 0 \\
0 & 0 & 2ic_i c_{i_z}
\end{pmatrix},
\end{equation}
where $P_0$ is a projection operator onto the 0-flux sector.

By summing up all the contributions from the three bonds surrounding
the $i$th site, the total EQM in the second order becomes
\begin{align}
    P_0 q_{i} P_0 =& \frac{t_2^2}{(U + \frac{3}{2}\lambda)^2}\begin{pmatrix}
3ic_i c_{i_x} & 0 & 0 \\
0 & 3ic_i c_{i_y} & 0 \\
0 & 0 & 3ic_i c_{i_z}
\end{pmatrix} \nonumber \\
&-\frac{t_2^2}{(U + \frac{3}{2}\lambda)^2}(ic_i c_{i_x}+ic_i c_{i_y}+ic_i c_{i_z})\openone_3,
\end{align}
which is nothing but a nematic order parameter as two terms cancel
out when $\langle ic_i c_{i_x}\rangle = \langle ic_i c_{i_y}\rangle = \langle ic_i c_{i_z}\rangle$
and the $C_3$ symmetry around the $i$th site is preserved.  Thus,
we have shown that EQM of Ru is directly connected to the nematic
order parameter of Majorana fermions.  Especially, a nematic Kitaev
spin liquid (NKSL) where the ground state remains the 0-flux sector
but breaks the $C_3$ symmetry by a nematic order
parameter can be detected through the measurement of this EQM
directly by Ru-NMR or Ru-MS.  However, such an effect could
compete with a static EQM coming from the trigonal distortion,
so we should be careful about whether $\eta$ is detectable if
we include both of the contributions.

\begin{figure}
    \centering
    \includegraphics[width=8.6cm]{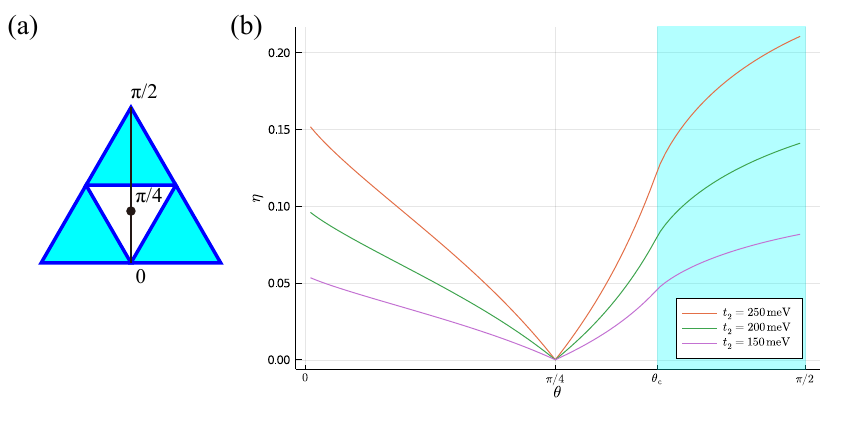}
    \caption{(a) Phase diagram of the Kitaev model~\cite{Kitaev2006}.
    Cyan regions represent $A$ phase, and a white region represents
    $B$ phase. A black solid line represents the $K^x=K^y$ line,
    which is parametrized by $\theta$ as depicted.
    (b) $\eta$ with respect to the model parameter $\theta$.
    $\Delta = 10$ meV, $\lambda = 150$ meV, and $U = 1.5$ eV are
    used.  $t_2$ takes 150, 200, and 250 meV.  Kitaev's gapped
    $A$ phase is shown by a cyan shaded region.}
    \label{eta}
\end{figure}    

\textit{Trigonal distortion}.---
Even if we introduce a small trigonal distortion $\Delta \neq 0$,
the ground state remains a Kramers doublet in the atomic limit and
the effective spin-1/2 description is valid.  The effective operator
form of EQM can be obtained almost in the same way as before
up to the first order in $\Delta/\lambda$.

\begin{align}
    P_0 q_{i} P_0 =& \begin{pmatrix}
\frac{3it_2^2}{(U + \frac{3}{2}\lambda)^2}c_i c_{i_x} & -\frac{4\Delta}{3\lambda} & -\frac{4\Delta}{3\lambda} \\
-\frac{4\Delta}{3\lambda} & \frac{3it_2^2}{(U + \frac{3}{2}\lambda)^2}c_i c_{i_y} & -\frac{4\Delta}{3\lambda} \\
-\frac{4\Delta}{3\lambda} & -\frac{4\Delta}{3\lambda} & \frac{3it_2^2}{(U + \frac{3}{2}\lambda)^2}c_i c_{i_z}
\end{pmatrix} \nonumber \\
&-\frac{t_2^2}{(U + \frac{3}{2}\lambda)^2}(ic_i c_{i_x}+ic_i c_{i_y}+ic_i c_{i_z})\openone_3 \nonumber \\
&+O(\Delta^2,\Delta t_2^2, t_2^4).
\end{align}
By diagonalizing this tensor, we can calculate the value of $\eta$.
Since usually $\Delta/\lambda > t_2^2/U^2$, the principal $a$-axis
is nearly perpendicular to the (111) plane.  $b$- and $c$-axes
are inside this plane, detecting the $C_3$ symmetry of the system.

In order to show the relevance of our theory to detect NKSL,
we try to check the size of $\eta$ for the ansatz state.
In the mean-field level, the ansatz state of NKSL should
be the ground state for the following ansatz Hamiltonian.
\begin{equation}
    \mathcal{H}_\textrm{NKSL}=-\sum_{\langle ij\rangle \in \gamma} K^\gamma S_i^\gamma S_j^\gamma,\label{NKSL}
\end{equation}
where $K^\gamma > 0$ is an effective Kitaev interaction for
the $\gamma$-direction.  On the $K^x = K^y$ line shown in
Fig.~\ref{eta}(a), Lieb's theorem~\cite{Lieb1994} is applicable
and the expectation value of EQM becomes
\begin{equation}
    \braket{\Psi_\textrm{GS}|q_i|\Psi_\textrm{GS}} = \braket{\Psi_\textrm{GS}|P_0 q_{i} P_0|\Psi_\textrm{GS}},
\end{equation}
for any ground state $\ket{\Psi_\textrm{GS}}$.  We then
compute $\eta$ for the ground state of $\mathcal{H}_\textrm{NKSL}$
along the line $K^x = K^y$.  The results are shown in Fig.~\ref{eta}(b),
where $\theta$ is defined as $\tan \theta = K^z/K^x$.
The calculation method is included in SM~\cite{SM}.

From the isotropic point $\theta=\pi/4$ with $\eta=0$, the value
of $\eta$ gradually grows, and continuously changes around
$\theta=\theta_\textrm{c}$ with $\tan \theta_\textrm{c} = 2$,
where the topological transition between Kitaev's $B$ and $A$
phases occurs.  In the gapped $A$ phase (cyan shaded region),
$\eta$ reaches 0.1--0.2.  Thus, the topological nematic
transition should result in $O(0.1)$ change of the value of $\eta$,
which is definitely detectable in the Ru-NMR or Ru-MS measurement.

Though the transition is continuous, the derivative
of $\eta$ has a cusp at the transition point (see Fig.~S2 in
SM~\cite{SM}).  Experimentally, the $B$ phase and the $A$ phase
can be distinguished by the presence of a cusp in the derivative,
and the critical value can be determined by its position.
The consequence of an applied magnetic field is also discussed
in SM~\cite{SM}.

\textit{Other contributions}.---
In this Letter, we have only considered the onsite $d$-orbital
contribution to EFG.  Usually, the interaction with EFG is divided
into onsite and offsite contributions~\cite{Abragam1970} as
$\mathcal{H}_\textrm{el} = \mathcal{H}_\textrm{el}^\textrm{on} + \mathcal{H}_\textrm{el}^\textrm{out}$ with
\begin{align}
    \mathcal{H}_\textrm{el}^\textrm{on} &=-\frac{e^2 Q}{2I(2I-1)}\langle r^{-3} \rangle \braket{L|\!|\alpha|\!|L} \bm{I}\tensor{q}\bm{I}, \nonumber \\
    \mathcal{H}_\textrm{el}^\textrm{out} &=(1-\gamma_\infty)\frac{eQ}{2I(2I-1)} \bm{I}\tensor{V}^\textrm{out}\bm{I},
\end{align}
where $e$ is the elementary charge, $Q$ is the quadrupole moment
of the nucleus, $\bm{I}$ are nuclear spin operators where $I$
depends on the isotope, $\langle r^{-3} \rangle$ is the expectation
value of $r^{-3}$ for Ru $4d$-electrons, $\braket{L|\!|\alpha|\!|L}$
is a constant defined in Ref.~\cite{Abragam1970},
$\gamma_\infty$ is the Sternheimer antishielding factor, and
$\tensor{V}^\textrm{out}$ is the EFG tensor caused by the
surrounding ions.

Usually, $\mathcal{H}_\textrm{el}^\textrm{on}$ is the main
contribution as Ru $4d$-orbitals are strongly localized, and thus
we have ignored the effect of $\mathcal{H}_\textrm{el}^\textrm{out}$
so far.  However, because the $C_3$ symmetric structure of ligands
is stable in $\alpha$-RuCl$_3$, the effect of $\tensor{V}^\textrm{out}$
is just renormalizing the value of $\Delta$.  Therefore, our theory
is qualitatively valid even if we include the contribution from
the surrounding ions.  Whether or not it gives a nonnegligible
change quantitatively will be discussed in the future.

\textit{Discussion}.---
We have shown that the nematic transition in $\alpha$-RuCl$_3$
is detectable by NMR and MS through the measurement of $\eta$.
Experiments should be combined with the high-resolution
X-ray diffraction to exclude the possibility of a lattice distortion.
While the conclusion is modified when the external magnetic field
is applied, the first-order contribution vanishes and $\eta$
still serves as a nematic order parameter.  The mechanism of the
detection itself is different from conventional electronic nematic
phases.  Although the expression of $q$ given by the
bilinear form of the spin operators is not limited to Kitaev systems,
its highly anisotropic form is a consequence of the strong SOC.

Our theory can be generalized to the three-dimensional
extensions of the Kitaev model~\cite{Obrien2016,Yamada2017XSL}.
Especially, the spin-Peierls instability expected in the hyperoctagon
lattice~\cite{Hermanns2015} is potentially detectable in our scheme
based on NMR and MS.

In the case of NMR, not only static quantities like EFG,
but also dynamical quantities can be observed.  Especially,
the nuclear spin-lattice relaxation rate divided by temperature
$1/T_1T$ would also be a good probe for the time scale of
the nematic transition.  We would remark that the anisotropy of
$1/T_1T$ can be another signature of the existence of a
nematic order~\cite{Smerald2016}.

\begin{acknowledgments}
We thank K.~Ishida, Y.~Matsuda, T.~Shibauchi, S.~Suetsugu, and
Y.~Tada for fruitful discussions.
This work was supported by the Grant-in-Aids for Scientific Research
from MEXT of Japan (Grant Nos. JP17K05517 and JP21H01039),
and JST CREST Grant Number JPMJCR19T5, Japan.
\end{acknowledgments}

\bibliography{paper}

\end{document}


\title{Supplemental Material for ``Electric probe for the toric code phase in Kitaev materials \\
through the hyperfine interaction''}

\author{Masahiko G. Yamada}
\email[]{myamada@mp.es.osaka-u.ac.jp}
\affiliation{Department of Materials Engineering Science, Osaka University, Toyonaka 560-8531, Japan}
\author{Satoshi Fujimoto}
\affiliation{Department of Materials Engineering Science, Osaka University, Toyonaka 560-8531, Japan}
\affiliation{Center for Quantum Information and Quantum Biology, Osaka University, Toyonaka 560-8531, Japan}

\maketitle

\onecolumngrid

\section{Contribution from the $e_g$-orbitals}

We quickly discuss the nonnegligible contribution from
the $e_g$-orbitals.  In this section, we ignore a spin index
for simplicity.  In the basis set for the $d$-orbitals with
\begin{equation}
    \bm{d}^\dagger = \left(d_{yz}^\dagger, d_{xz}^\dagger, d_{xy}^\dagger, d_{3z^2-r^2}^\dagger, d_{x^2-y^2}^\dagger \right),
\end{equation}
the angular momentum operator $\bm{L}$ can be written as
follows:
\begin{equation}
L^x = \bm{d}^\dagger \begin{pmatrix}
    0 & 0 & 0 & -\sqrt{3}i & -i \\
    0 & 0 & i & 0 & 0 \\
    0 & -i & 0 & 0 & 0 \\
    \sqrt{3}i & 0 & 0 & 0 & 0 \\
    i & 0 & 0 & 0 & 0
\end{pmatrix}\bm{d},\quad
L^y = \bm{d}^\dagger \begin{pmatrix}
    0 & 0 & -i & 0 & 0 \\
    0 & 0 & 0 & \sqrt{3}i & -i \\
    i & 0 & 0 & 0 & 0 \\
    0 & -\sqrt{3}i & 0 & 0 & 0 \\
    0 & i & 0 & 0 & 0
\end{pmatrix}\bm{d},\quad
L^z = \bm{d}^\dagger \begin{pmatrix}
    0 & i & 0 & 0 & 0 \\
    -i & 0 & 0 & 0 & 0 \\
    0 & 0 & 0 & 0 & 2i \\
    0 & 0 & 0 & 0 & 0 \\
    0 & 0 & -2i & 0 & 0
\end{pmatrix}\bm{d}.
\end{equation}
Though the up-left $3 \times 3$ part coincides with $-\bm{l}_\textrm{eff}$,
there are offdiagonal elements in $\bm{L}$~\cite{Stamokostas2018}.
Thus, the product of the two has a nontrivial contribution from
the $e_g$-orbitals.

\section{Computation of the anisotropy parameter}

In a nematic Kitaev spin liquid (NKSL),
the anisotropy parameter $\eta$ can be computed as follows.
First, we consider the case without a magnetic field.
In this case, the Hamiltonian is $\mathcal{H}_\textrm{NKSL}$
in the main text and the ground state is exactly solvable
by introducing Majorana fermions~\cite{Kitaev2006}.
\begin{equation}
    \ket{\Psi_\textrm{GS}}=P_\textrm{phys}\ket{\Psi},
\end{equation}
where $P_\textrm{phys}$ is a projection onto the physical
subspace, and $\ket{\Psi}$ is the ground state of the itinerant
Majorana fermions $c_i$.  Thus,
\begin{equation}
    \langle ic_i c_j \rangle = \braket{\Psi_\textrm{GS}|ic_i c_j|\Psi_\textrm{GS}} = \braket{\Psi|P_\textrm{phys} ic_i c_j P_\textrm{phys}|\Psi}
\end{equation}

Meanwhile, $\mathcal{H}_\textrm{NKSL}$ obeys
\begin{equation}
    \braket{\Psi_\textrm{GS}|\mathcal{H}_\textrm{NKSL}|\Psi_\textrm{GS}}
    =\sum_{\langle ij\rangle \in \gamma} \frac{K^\gamma}{4} \braket{\Psi|P_\textrm{phys} ic_i c_j P_\textrm{phys}|\Psi},
\end{equation}
assuming that the ground state is in the 0-flux sector.
Thus, from the Hellmann-Feynman theorem $\langle ic_i c_j \rangle$
for the $\gamma$-direction can be computed by
\begin{equation}
    \langle ic_i c_j \rangle_\gamma = \frac{4}{N_\textrm{unit}} \frac{\partial}{\partial K^\gamma} \braket{\Psi_\textrm{GS}|\mathcal{H}_\textrm{NKSL}|\Psi_\textrm{GS}},
\end{equation}
where $N_\textrm{unit}$ is the number of unit cells~\cite{Li2018}.
We here note that in the thermodynamic limit physical states
and unphysical states have the same energy and the same expectation
value for physical observables~\cite{Zschocke2015}.  Thus, we can
ignore $P_\textrm{phys}$ and use $\ket{\Psi}$ directly to calculate
observables, as long as the thermodynamic limit can be taken
analytically.

\begin{figure}
    \centering
    \includegraphics[width=8.6cm]{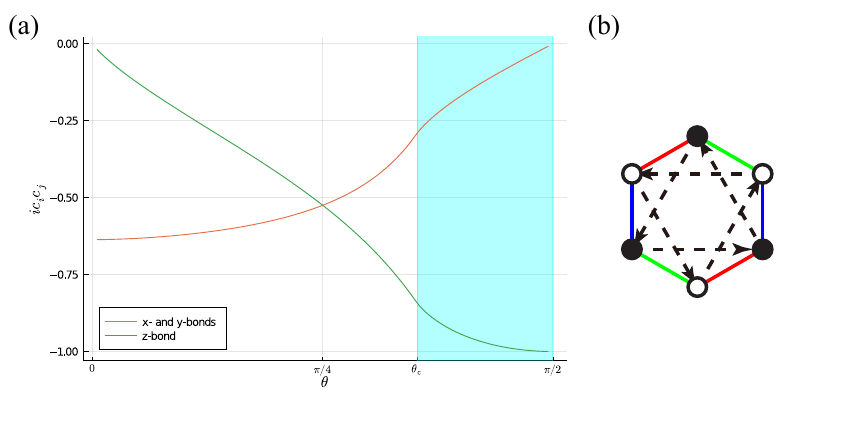}
    \caption{(a) $\langle ic_i c_j \rangle$ for each bond direction.
    The definition of $\theta$ is included in the main text.
    A cyan shaded region represents Kitaev's $A$ phase.
    (b) Honeycomb lattice with a direction of the NNN bonds
    shown in black dashed arrows.}
    \label{icicj}
\end{figure}

\begin{figure}
    \centering
    \includegraphics[width=8.6cm]{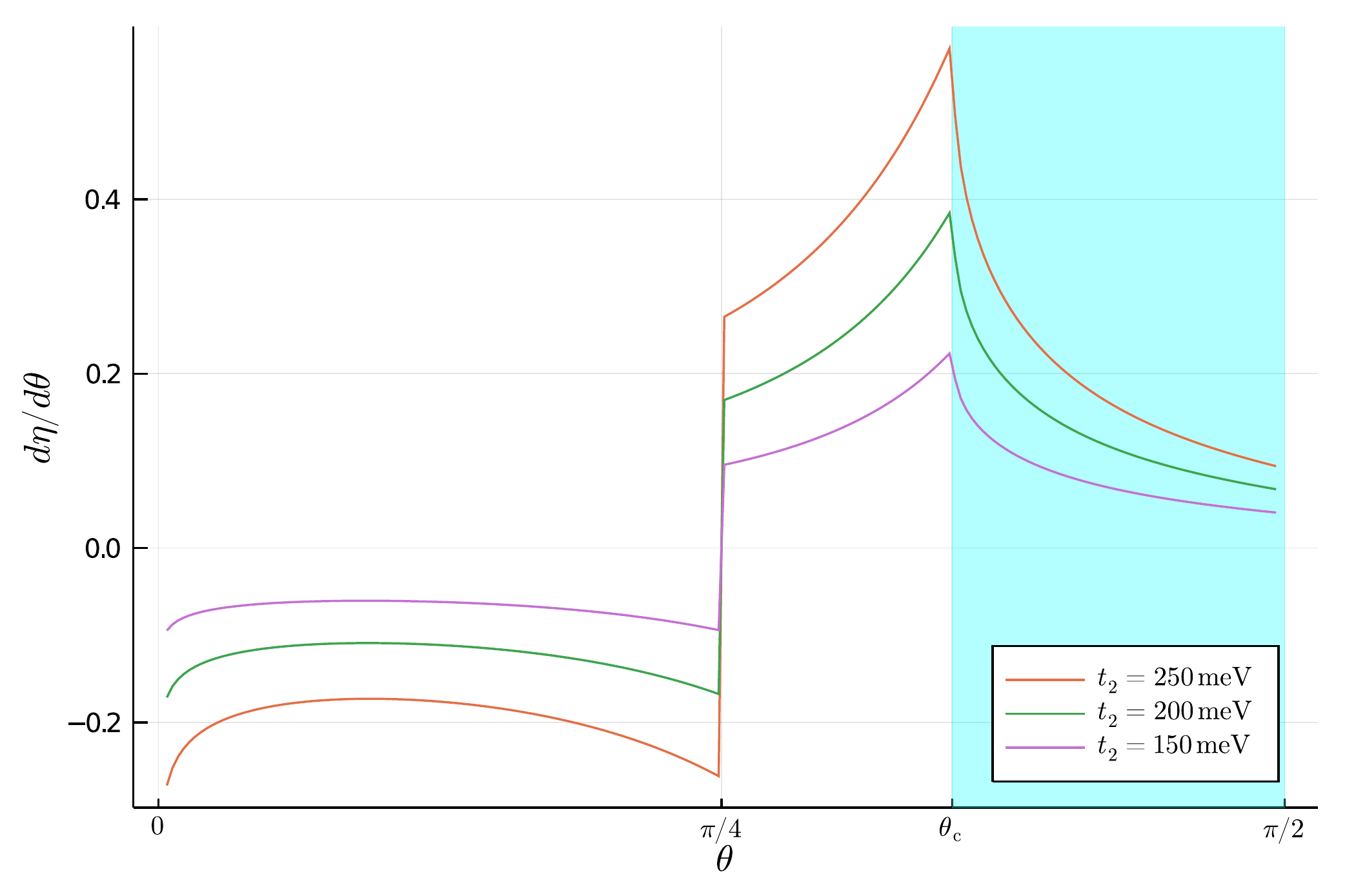}
    \caption{Derivative of $\eta$. A cyan shaded region represents
    Kitaev's $A$ phase.  As clearly seen, $d\eta/d\theta$ has
    a cusp at the transition point $\theta = \theta_\textrm{c}$
    between the $B$ phase and the $A$ phase.}
    \label{derivative}
\end{figure}

The direct computation gives
\begin{equation}
    E = \braket{\Psi|\mathcal{H}_\textrm{NKSL}|\Psi} = -\frac{1}{16\pi^2}\int_0^{2\pi}dk_1 \int_0^{2\pi}dk_2 \varepsilon_k,
\end{equation}
where $\varepsilon_k = |K^x e^{ik_1} + K^y e^{ik_2} + K^z|$.
Thus,
\begin{align}
    \langle ic_i c_j \rangle_x &= -\frac{1}{4\pi^2}\int_\textrm{BZ} d^2k \frac{(K^x \cos k_1 + K^y \cos k_2 + K^z)\cos k_1 + (K^x \sin k_1 + K^y \sin k_2)\sin k_1}{\varepsilon_k},\\
    \langle ic_i c_j \rangle_y &= -\frac{1}{4\pi^2}\int_\textrm{BZ} d^2k \frac{(K^x \cos k_1 + K^y \cos k_2 + K^z)\cos k_2 + (K^x \sin k_1 + K^y \sin k_2)\sin k_2}{\varepsilon_k},\\
    \langle ic_i c_j \rangle_z &= -\frac{1}{4\pi^2}\int_\textrm{BZ} d^2k \frac{K^x \cos k_1 + K^y \cos k_2 + K^z}{\varepsilon_k},
\end{align}
where the numerical integration is done in the whole
Brillouin zone ($\textrm{BZ}$), $0 \leq k_1 \leq 2\pi$ and
$0 \leq k_2 \leq 2\pi$.  The expectation values of $ic_i c_j$
are plotted in Fig.~\ref{icicj}(a).  We also plot the derivative
of $\eta$ in the same strategy as shown in Fig.~\ref{derivative}.

The same strategy applies to the case without a time-reversal
symmetry.  For simplicity, we only consider the case where
$K^x=K^y=K^z=K$.  The Hamiltonian considered here is
\begin{equation}
    \mathcal{H}_\textrm{mag} = -K\sum_{\langle ij\rangle \in \gamma} S_i^\gamma S_j^\gamma-\kappa \sum_{\langle ijk \rangle_{\alpha\beta\gamma}} S_i^\alpha S_j^\beta S_k^\gamma,
\end{equation}
where $\kappa$ is a real time-reversal breaking parameter.
$\langle ijk \rangle_{\alpha\beta\gamma}$ means nearest-neighbor (NN)
pairs, where $\langle ij \rangle$ and $\langle jk \rangle$ are on the
$\alpha$- and $\gamma$-directions, respectively, and $\beta$ is the
component which is neither $\alpha$ nor $\gamma$.

As is well-known, the $\kappa$ term is nothing but a
next-nearest-neighbor (NNN) hopping term for Majorana fermions.
Thus, we can compute the expectation value of NNN hopping terms
from the derivative about $\kappa$.  The final expression becomes
\begin{align}
    \langle ic_i c_j \rangle_\textrm{NN} &= -\frac{1}{12\pi^2}\int_\textrm{BZ} d^2k \frac{K|e^{ik_1}+e^{ik_2}+1|^2}{\epsilon_k}, \\
    \langle ic_i c_j \rangle_\textrm{NNN} &= -\frac{1}{12\pi^2}\int_\textrm{BZ} d^2k \frac{\kappa[\sin k_1-\sin k_2 + \sin(k_2-k_1)]^2}{\epsilon_k},
\end{align}
where $\langle ic_i c_j \rangle_\textrm{NN}$ is a bond expectation value
for NN bonds, $\langle ic_i c_j \rangle_\textrm{NNN}$ is a bond
expectation value for NNN bonds, and
$\epsilon_k = \sqrt{K^2|e^{ik_1}+e^{ik_2}+1|^2 + \kappa^2[\sin k_1-\sin k_2 + \sin(k_2-k_1)]^2}$.
As for NNN bonds, the bond direction is defined as illustrated in
Fig.~\ref{icicj}(b), where the arrow points from $j$ to $i$.
These formulae will be used in the next section.

\section{Magnetic field effect}.

In the main text, we have assumed the existence of a time-reversal
symmetry, but in real NMR or MS experiments an external magnetic field
is usually applied.  First, we modify the ansatz Hamiltonian as follows:
\begin{equation}
    \mathcal{H}_\textrm{Zeeman}=-\sum_{\langle ij\rangle \in \gamma} K^\gamma S_i^\gamma S_j^\gamma-\bm{h}\cdot\sum_i\bm{S}_i,
\end{equation}
where $\bm{h}=(h^x, h^y, h^z)$ represents the normalized external field.

In the first-order in $\bm{h}$, the wavefunction can be approximated as
\begin{equation}
    \ket{\Psi^{(1)}}\sim\ket{\Psi_\textrm{GS}}-\frac{1-P_0}{-\Delta_\textrm{flux}}\bm{h}\cdot\sum_i\bm{S}_i\ket{\Psi_\textrm{GS}},
\end{equation}
where $\Delta_\textrm{flux}\sim 0.26K/4$ is a flux gap energy for
two neighboring vortices.  In the first order, only the
contribution that creates or annihilates two neighboring
vortices survives.  This condition holds for $S_i^x S_{i_z}^z$
and $S_i^y S_{i_z}^z$ in Eq.~(12) in the main text, for example,
which results in the following NNN hopping operators,
after summing up three surrounding bonds around the $i$th site:
\begin{align}
&P_0 q_{i} P_0 = \nonumber \\
&\begin{pmatrix}
\frac{3it_2^2}{(U + \frac{3}{2}\lambda)^2}c_i c_{i_x} & -\frac{4\Delta}{3\lambda} + \frac{4it_2^2 h^z}{3(U + \frac{3}{2}\lambda)^2\Delta_\textrm{flux}}(c_{i_z}c_{i_x} + c_{i_y}c_{i_z}) & -\frac{4\Delta}{3\lambda} + \frac{4it_2^2 h^y}{3(U + \frac{3}{2}\lambda)^2\Delta_\textrm{flux}}(c_{i_y}c_{i_z} + c_{i_x}c_{i_y}) \\
-\frac{4\Delta}{3\lambda} + \frac{4it_2^2 h^z}{3(U + \frac{3}{2}\lambda)^2\Delta_\textrm{flux}}(c_{i_z}c_{i_x} + c_{i_y}c_{i_z}) & \frac{3it_2^2}{(U + \frac{3}{2}\lambda)^2}c_i c_{i_y} & -\frac{4\Delta}{3\lambda} + \frac{4it_2^2 h^x}{3(U + \frac{3}{2}\lambda)^2\Delta_\textrm{flux}}(c_{i_z}c_{i_x} + c_{i_x}c_{i_y}) \\
-\frac{4\Delta}{3\lambda} + \frac{4it_2^2 h^y}{3(U + \frac{3}{2}\lambda)^2\Delta_\textrm{flux}}(c_{i_y}c_{i_z} + c_{i_x}c_{i_y}) & -\frac{4\Delta}{3\lambda} + \frac{4it_2^2 h^x}{3(U + \frac{3}{2}\lambda)^2\Delta_\textrm{flux}}(c_{i_z}c_{i_x} + c_{i_x}c_{i_y}) & \frac{3it_2^2}{(U + \frac{3}{2}\lambda)^2}c_i c_{i_z}
\end{pmatrix} \nonumber \\
&-\frac{t_2^2}{(U + \frac{3}{2}\lambda)^2}(ic_i c_{i_x}+ic_i c_{i_y}+ic_i c_{i_z})\openone_3.
\end{align}

All new terms are NNN hopping operators breaking the time-reversal
symmetry, so their contribution vanishes in the computation of
$\braket{\Psi_\textrm{GS}|q_{i}|\Psi_\textrm{GS}}=\braket{\Psi_\textrm{GS}|P_0 q_{i} P_0|\Psi_\textrm{GS}}$,
as long as $\ket{\Psi_\textrm{GS}}$ is time-reversal symmetric.
Thus, the magnetic field effect is negligible in the first order.

Though we have shown that the time-reversal breaking effect is
negligible in the first-order perturbation, there could be
another effect which produces a nonzero expectation value for
NNN bonds.  For example, if the $\Gamma^\prime$ term coming
from the trigonal distortion is included in the magnetic Hamiltonian,
the gap opening by the time-reversal breaking can occur in
the first-order in $\bm{h}$~\cite{Takikawa2019,Takikawa2020}.
This type of effects can be included by adding a so-called
$\kappa$ term to the Hamiltonian by hand.
\begin{equation}
    \mathcal{H}_\textrm{eff}=-\sum_{\langle ij\rangle \in \gamma} K^\gamma S_i^\gamma S_j^\gamma-\kappa \sum_{\langle ijk \rangle_{\alpha\beta\gamma}} S_i^\alpha S_j^\beta S_k^\gamma -\bm{h}\cdot\sum_i\bm{S}_i.
\end{equation}
The $\kappa$ term explicitly breaks
the time-reversal symmetry and $\langle ic_i c_j \rangle$ for NNN
bonds no longer vanish~\footnote{It is not trivial to show that
$\kappa$ does not depend on the bond direction even if $K^\gamma$
are not equal. See \textit{e.g.} Eq.~(9) in Ref.~\cite{Yamada2020}.}.
Again we use the first-order perturbation in $\bm{h}$ and now the
magnetic field effect becomes nontrivial.  Even in the case where
$K^x=K^y=K^z$ without a nematic order, $\eta$ does not always vanish
unless the magnetic field is perpendicular to the (111) plane.
A careful treatment is necessary to detect a nematic order through
$\eta$.

\begin{figure}
    \centering
    \includegraphics[width=8.6cm]{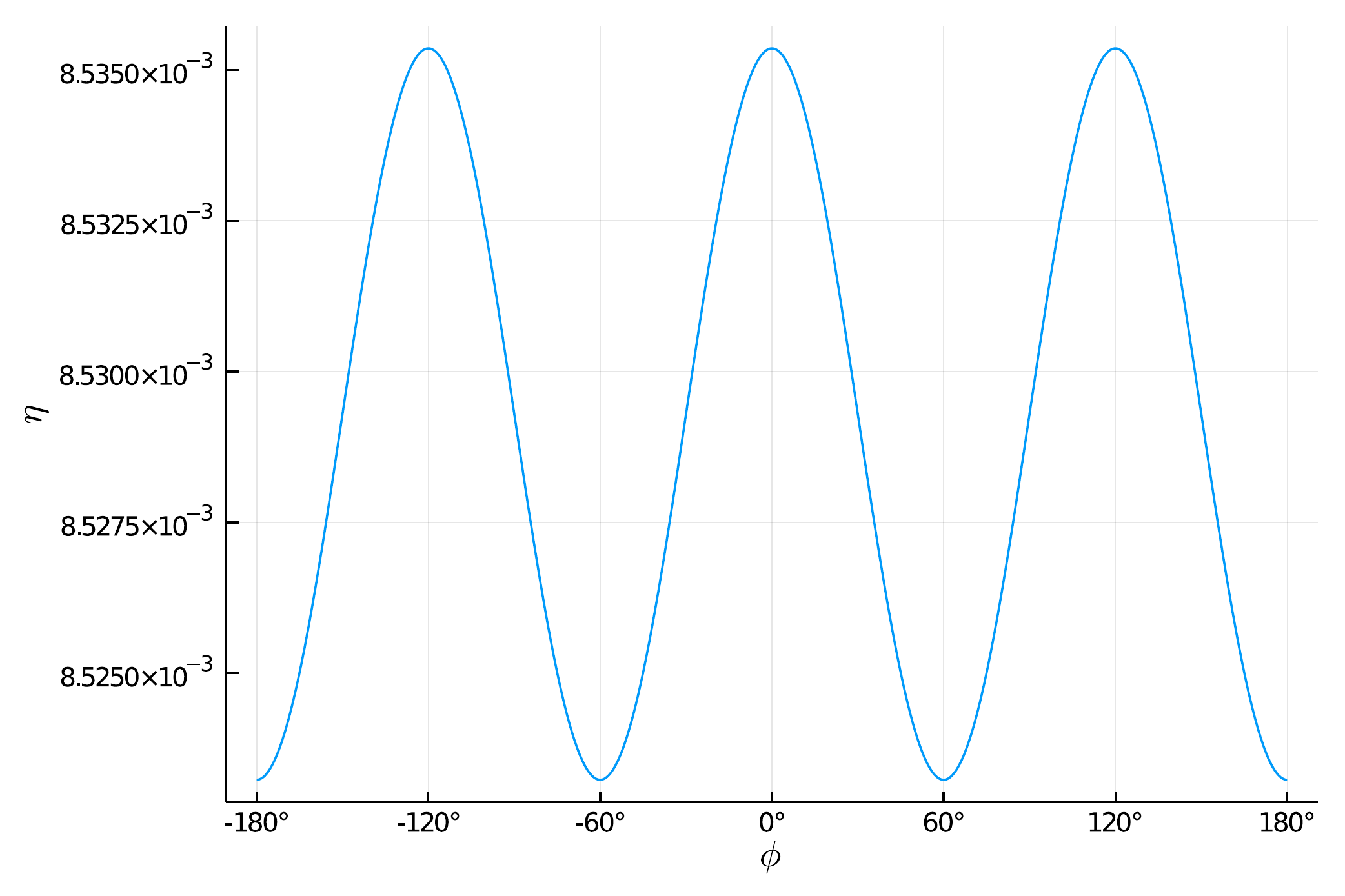}
    \caption{Field angle dependence of $\eta$.  The magnetic field
    is applied in the (111) plane with an angle $\phi$.  A typical
    ansatz of $\kappa/K=0.1$ and $|\bm{h}|/K=0.1$ is used.
    $\Delta = 10$ meV, $\lambda = 150$ meV, $t_2 = 200$ meV,
    and $U = 1.5$ eV are used.}
    \label{phi}
\end{figure}

Fortunately, however, this effect is very small when we apply an
in-plane magnetic field.  This is because the first-order contribution
of $\bm{h}$ for $\kappa(h^x, h^y, h^z)$ vanishes when $h^x+h^y+h^z=0$,
assuming the complete $C_3$ symmetry of the original electronic
Hamiltonian.  $\kappa$ is effectively expanded from the third order
in $\bm{h}$, and we can assume that $\kappa$ to be $O(h^3)$, which
is consistent with experiments~\cite{Tanaka2020}.

We plot the angle dependence of $\eta$ with respect to an azimuth
angle $\phi$ of the applied magnetic field in Fig.~\ref{phi} in
the isotropic case $K^x=K^y=K^z=K$.  As is clearly seen,
a small but nonzero $\eta$ appears due to the applied magnetic field.
It has a tiny angle dependence with a period of $120\degree$.
Thus, we can say that experiments should be done by comparing
$\phi = 0\degree$ with $120\degree$ to avoid this field effect.

To sum up, in reality $\eta$ is not exactly 0 even in the isotropic
case, so we should be careful about the actual experimental setup
to determine the critical point of the nematic transition.
Experiments should be combined with a measurement of an
anisotropy of the relaxation rate $1/T_1 T$, which is compared
between $\phi = 0\degree$ and $120\degree$, for example.

\bibliography{suppl}